# DIRECT STRAIN AND SLOPE MEASUREMENT USING 3D DSPSI

Wajdi Dandach[a], Jerome Molimard[b] and Pascal Picart[c]

[a]LTDS, UMR 5513, École Nationale Supérieure des Mines, SMS-EMSE, CNRS, Saint-Étienne, France
[b]LCG, UMR CNRS 5146, École Nationale Supérieure des Mines, CIS-EMSE, CNRS, Saint-Étienne, France
[c]LAUM, UMR CNRS 6613, ENSIM, Université du Maine, Rue Aristote - 72085 Le Mans cedex 09, France

dandach@emse.fr, molimard@emse.fr, pascal.picart@univ-lemans.fr

**ABSTRACT:**

this communication presents a new implementation of DSPSI. Its main features are 1. an advanced model taking into account the beam divergence, 2. the coupling with a surface shape measurement in order to generalize DSPSI to non-planar surfaces 3. the use of small shear distance made possible using a precise measurement procedure. A first application on a modified Iosipescu shear test is presented and compared to classical DIC measurements.

## 1. INTRODUCTION

Most of OFFT measure displacements (Electronic Speckle Pattern Interferometry, Grid Method, Moiré Interferometry and so on); one of the challenging parts of displacement measurement techniques is the calculation of the strain map from the displacement map. In particular, noise propagation and lens distortion has to be carefully treated. Adjacent to the methods measuring displacements, Shearography measures directly displacement derivatives of surfaces. More precisely, it eliminates the reference beam of holographic or speckle interferometry, which leads to a simplified optical set-up, not requiring special vibration isolation. These advantages have exhibited practically shearography a surface strain and rotation measurement system. A new implementation of shearography is presented here; its main features are 1. an advanced model taking into account the beam divergence, 2. the coupling with a surface shape measurement in order to generalize shearography to non-planar surfaces 3. the use of small shear distance made possible using a precise measurement procedure. Complete 3D feature is not presented here anyway, and a first application on a modified Iosipescu shear test is presented and compared to classical DIC measurements.

## 2. 2D/3D SHEAROGRAPHY SYSTEM

### 2.1. Description of shearography set-up

The set-up main architecture is classical, but light is conducted using optical fibers, as presented previously in [1] (Figure 1). A tunable laser illuminates the front surface of the specimen. Light is sequentially injected in the four optical fibers by using an optical switch. The outputs fibers are attached to a device, manufactured in the laboratory, which allows the laser beam illuminating the surface of the specimen by four directions, with equivalent illumination angle. Due to the global geometry anyway, the beam are uncollimated, and illumination angle (and consequently the sensitivity vectors) are not constant over the map. The diffused beam from the specimen is sheared by Michelson interferometer. First mirror is fixed; the second is controlled by a 3-axis PTZ device PSH 1z NV from Piezo-system Jena, capable of tilting or translating a mirror. Shearing in x or y directions is obtained by tilting one mirror, similarly phase stepping is realized with a piston movement of the mirror.

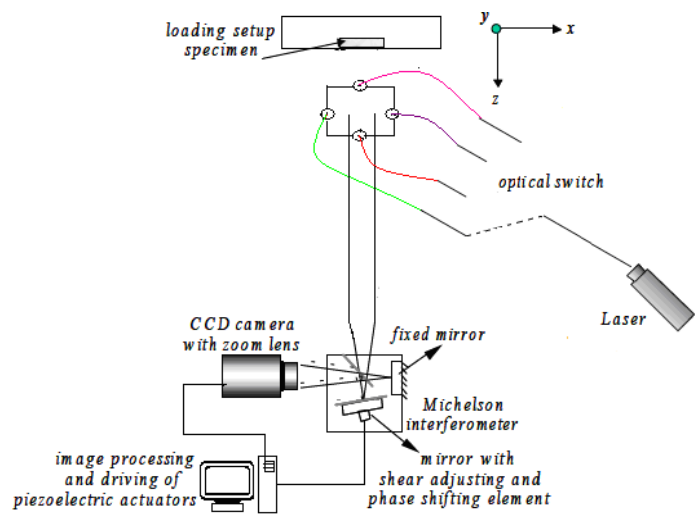

**Figure 1. Sherarography set-up**

Acquisition is performed using a 12 bit Jai camera equipped with a zoom lens, connected to a gigabit Ethernet card plugged into a PC. Images are grabbed by a Labview program that controls shear distance and phase stepping. Reference and current state are processed by a Matlab program (Photomecanix toolbox).

## 2.2. Extracting displacement gradients on non flat surfaces

The extraction of strains on a non flat surface implies the knowledge of the surface shape, the use of an optical model with local sensitivity vector, and a projection procedure of strain gradients onto the surface to derive strain. This complete approach being too heavy, it has been decided to focus on the optical model; details of 3D features (surface measurement, strain projection) will be given in another context.

The Michelson interferometer combine informations of points $P$ and $P_s$. The shearing distance between these two points can be noted $\vec{\delta x} = \begin{pmatrix} \delta x \\ \delta y \\ 0 \end{pmatrix}$.

$\vec{k}_x$ and $\vec{k}_o$ being the illumination and the observation vectors, the phase variation in the direction $\vec{\delta x}$ between two consecutive states writes:

$$\Delta \varphi = \frac{2\pi}{\lambda}(\vec{k}_x - \vec{k}_o) \cdot (\vec{\delta x}^2 - \vec{\delta x}^1) = \frac{2\pi}{\lambda}(\vec{k}_x - \vec{k}_o) \cdot \vec{\delta d}(P) \quad (1)$$

Because of the optical arrangement and because of the shape of the object, each point $M$ of the surface has a different sensitivity. For a given source $S_i$, with camera focus $C$, equation (1) becomes [2]:

$$\begin{aligned} \Delta \varphi_i^k &= \frac{2\pi}{\lambda}(\vec{k}_i - \vec{k}_0) \cdot \vec{\delta d}^k(P) \\ &= \frac{2\pi}{\lambda}\left( \frac{\vec{S_i M}}{\|\vec{S_i M}\|} - \frac{\vec{MC}}{\|\vec{MC}\|} \right) \cdot \vec{\delta d}^k(P) \\ &= (S_{1i} \quad S_{2i} \quad S_{3i}) \cdot \vec{\delta d}^k(P) \end{aligned} \quad (2)$$

$\vec{\delta d}(P)$ is the differential displacement along $\vec{\delta x}$ direction, such as:

$$\vec{\delta d}^k(P) = \begin{bmatrix} \frac{\partial u}{\partial x} & \frac{\partial u}{\partial y} & \frac{\partial u}{\partial z} \\ \frac{\partial v}{\partial x} & \frac{\partial v}{\partial y} & \frac{\partial v}{\partial z} \\ \frac{\partial w}{\partial x} & \frac{\partial w}{\partial y} & \frac{\partial w}{\partial z} \end{bmatrix} \cdot \begin{pmatrix} \delta x^k \\ \delta y^k \\ 0 \end{pmatrix} \quad (3)$$

Note that no shear can be applied to the z direction. If taking into account the 4 illuminations and the two shearing directions, (2) and (3) give:

$$\begin{bmatrix} \Delta \varphi_1^1 & \Delta \varphi_1^2 \\ \Delta \varphi_2^1 & \Delta \varphi_2^2 \\ \Delta \varphi_3^1 & \Delta \varphi_3^2 \\ \Delta \varphi_4^1 & \Delta \varphi_4^2 \end{bmatrix} = \begin{pmatrix} S_{11} & S_{21} & S_{31} \\ S_{12} & S_{22} & S_{32} \\ S_{13} & S_{23} & S_{33} \\ S_{14} & S_{24} & S_{34} \end{pmatrix} \begin{bmatrix} \frac{\partial u}{\partial x} & \frac{\partial u}{\partial y} \\ \frac{\partial v}{\partial x} & \frac{\partial v}{\partial y} \\ \frac{\partial w}{\partial x} & \frac{\partial w}{\partial y} \end{bmatrix} \cdot \begin{pmatrix} \delta x^1 & \delta x^2 \\ \delta y^1 & \delta y^2 \end{pmatrix} \quad (4)$$

or:

$$[\Delta \varphi] = [S] \cdot [grad\, \vec{u}] \cdot [\delta_x] \quad (5)$$

Finally,

$$[grad\, \vec{u}] = ({}^t[S][S])^{-1}\, {}^t[S] \cdot [\Delta \varphi] \cdot ({}^t[\delta_x][\delta_x])^{-1}\, {}^t[\delta_x] \quad (6)$$

Equation (6) takes advantage of the information of the 4 beams implicitly using a least square formulation, thus increasing the robustness of the approach.

For a flat surface of an object in the (*x,y*) global frame of reference, strain and surface rotations are directly derived using their definitions:

$$\epsilon_{xx}=\frac{\partial u}{\partial x} \quad , \quad \epsilon_{yy}=\frac{\partial v}{\partial y} \quad , \quad \epsilon_{xy}=\frac{1}{2}\left(\frac{\partial u}{\partial y}+\frac{\partial v}{\partial x}\right) \quad , \quad \omega=\frac{1}{2}\left(\frac{\partial u}{\partial y}-\frac{\partial v}{\partial x}\right) \quad \text{and} \quad \frac{\partial w}{\partial x} \quad , \quad \frac{\partial w}{\partial y} \quad .$$

## 3. 1ST EXPERIMENTAL RESULTS

### 3.1. Mechanical set-up

The Iosipescu mechanical set-up based on the EMSE fixture [3] is presented in Fig 2. Measurements are performed in the middle of the specimen (grid lines). On the movable jaw we imposed a vertical displacement. A classical load cell is used to control the load. The specimen has been submitted to different loads from 0 to 400 N, with steps of 100 N. The sample width is 20 mm and its thickness 3 mm, the distance between the two clamps being 17.9 mm; its material is a Polyvinyl chloride. Last, pixel size in the object plane is 19.6 µm.

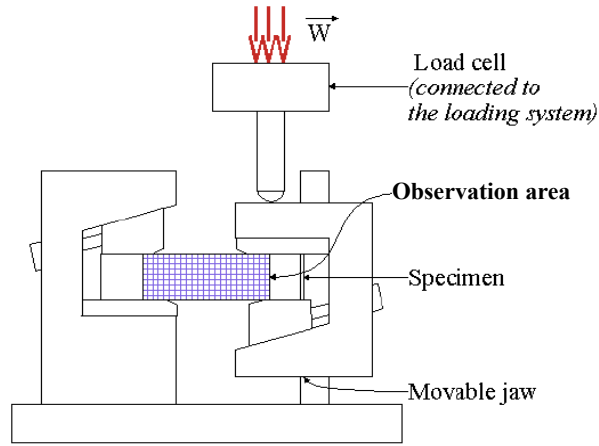

**Figure 2: Iosipescu Mechanical set-up**

### 3.2. Shearography / DIC Comparison

DIC is a home-made frequency-based implementation described in [4]. Derivation is based on a 2D polynomial approximation (2nd order, 9×9 kernel). Metrological data on the displacement and strain are given in table 1. Results of figure 3 show that the shearography system is more sensitive, even with a smaller spatial resolution. Of course, total implementation is more complicated, but automation makes acquisition and treatment easier than ever.

|  | DIC displacement | DIC strain | DSPSI strain |
|---|---|---|---|
| Resolution (σ) | 0.04 pix | 1000 µm/m | 2.3 µm/m |
| Spatial resolution (autocorrelation with 50 % cut-off) | 2.2 pix | 11.3 pix | 3.4 pix |

Table 1: metrological characteristics

## 4. CONCLUSIONS

A general formulation of DSPSI is proposed, compatible to measurements on a non planar surface. The approach is more stable and more precise than previous ones by taking advantage of the 4 illuminations. A first example on Iosipescu shear test shows results qualitatively better than a classical DIC implementation.

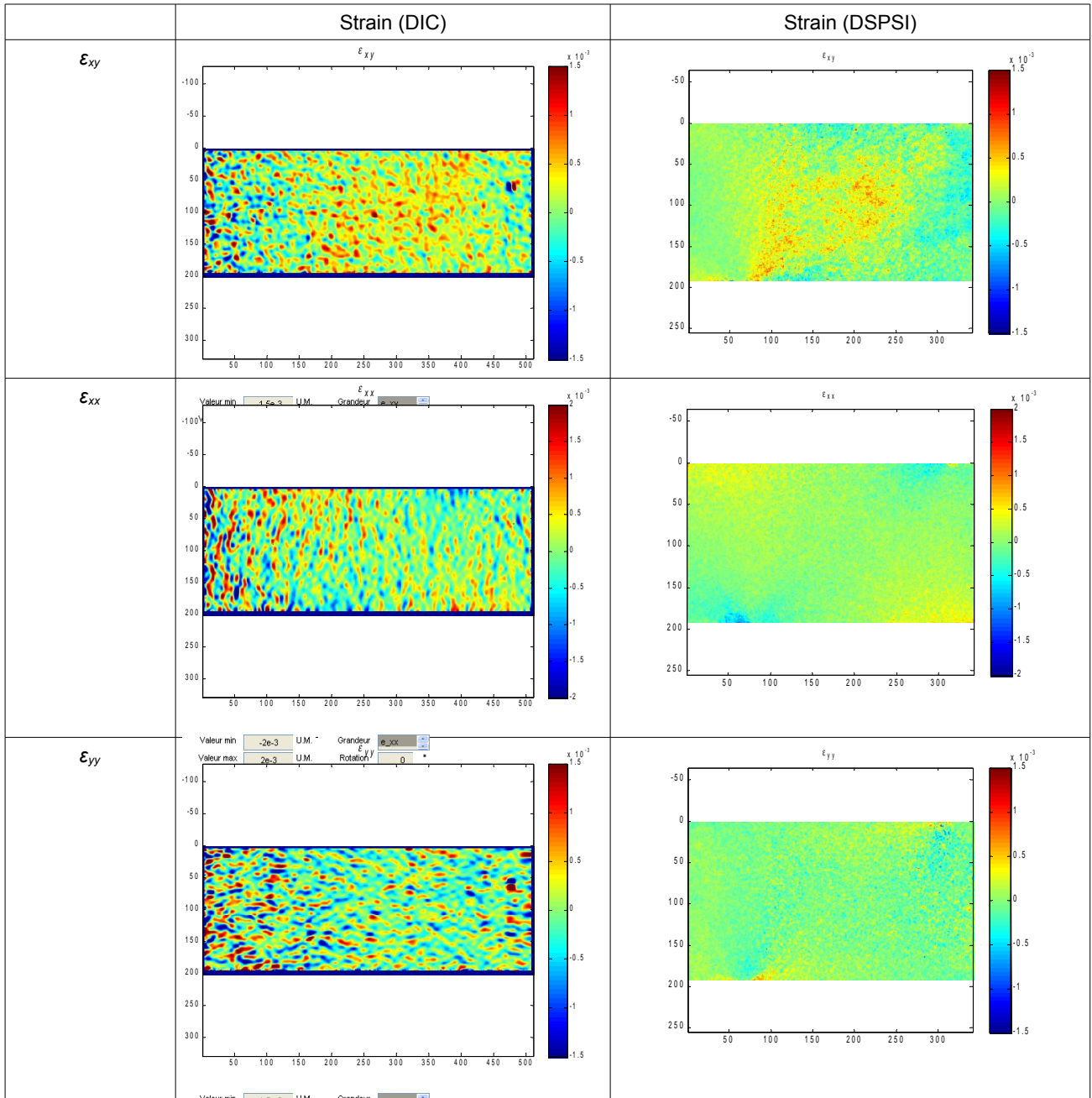

**Figure 3: Strain maps from Iosipescu shear test. DIC and DSPSI comparison**

*This work has been supported by French grant program ANR "Jeune Chercheur" 2009*

## 5. REFERENCES

[1] J. Molimard, D. Bounda, A. Vautrin, Quantitative strain and slope evaluation on a double lap joint tensile test using ESPSI, (2006) Proceedings Vol. 6341 Speckle06: Speckles, From Grains to Flowers, Pierre Slangen; Christine Cerruti, Editors, 63412R.

[2] J. Molimard, A. Dolinko, G. Kaufmann, Experimental study of thick composites stability under thermal loading using 3D ESPI set-up, (2009) in Optical Measurement Techniques for Systems & Structures G. J. Dirckx, J. B. (Ed.), Shaker Publishing, 255-264

[3] Pierron, F., design and experimental procedure, Internal Report n° 940125, Ecole des Mines de Saint-Etienne (France), 1994.

[4] Molimard, J.; Boyer, G. & H., Z. Frequency-based image analysis of random patterns: an alternative way to classical stereocorrelation Journal of the Korean Society of Non Destructive Testing, 2010, 30, 181-193